\documentclass[sigconf]{acmart}

\AtBeginDocument{%
  }

\usepackage{tcolorbox}
\usepackage{tikz}
\usepackage{listings}
\usepackage{subcaption}
\usepackage{xcolor}
\usepackage{makecell} 
\usepackage{tabularx}
\usepackage{multirow}
\usepackage{pgfplots}
\pgfplotsset{compat=1.18}
\usepackage{pgfplotstable}
\usepackage{xspace}
\usepackage{todonotes}
\usepackage{float}
\usepackage{graphicx}
\usepackage{booktabs}
\usepackage{bm}
\usepackage[normalem]{ulem}
\usepackage{xurl}

\usepackage{hyperref}
\usepackage{threeparttable} 
\usepackage{amsmath}
\usepackage{cleveref}
\usepackage{arydshln}
\usepackage{newtxtext}
\usepackage{newtxmath}

\usepackage{graphicx}
\usepackage{adjustbox}

\usepackage{tikz}
\usetikzlibrary{arrows.meta, positioning}

\crefformat{section}{\S#2#1#3}
\crefname{figure}{Figure}{Figures}
\crefname{table}{Table}{Tables}
\crefname{listing}{Listing}{Listings}
\crefname{theorem}{Theorem}{Theorems}
\crefname{thm}{Theorem}{Theorems}
\crefname{lemma}{Lemma}{Lemmata}
\crefname{equation}{Eqt.}{Eqts.}

\tcbset{
    mysummarybox/.style={
        colframe=black!50,
        colback=gray!10,
        boxrule=0.5mm,
        arc=1mm,
        top=0.5mm,
        bottom=0.5mm,
        left=0.5mm,
        right=0.5mm
    }
}

\newcommand{\blade}{{\textsc{Blade}}\xspace}
\newcommand{\chisel}{{\textsc{Chisel}}\xspace}
\newcommand{\razor}{{\textsc{Razor}}\xspace}
\newcommand{\occam}{{\textsc{Occam}}\xspace}
\newcommand{\trimmer}{{\textsc{Trimmer}}\xspace}
\newcommand{\lmcas}{{\textsc{Lmcas}}\xspace}
\newcommand{\cov}{{\textsc{Cov}}\xspace}
\newcommand{\cova}{{\textsc{Cov$_{A}$}}\xspace}

\newcommand{\chiselbench}{{\textsc{ChiselBench}}\xspace}
\newcommand{\sir}{\textsc{SIR}\xspace}

\newcommand{\napps}{11\xspace}

\newcommand{\mkdirutil}{\texttt{mkdir-5.2.1}\xspace}
\newcommand{\sortutil}{\texttt{sort-8.16}\xspace}
\newcommand{\uniqutil}{\texttt{uniq-8.16}\xspace}
\newcommand{\rmutil}{\texttt{rm-8.4}\xspace}
\newcommand{\bziputil}{\texttt{bzip2-1.0.5}\xspace}
\newcommand{\chownutil}{\texttt{chown-8.2}\xspace}
\newcommand{\dateutil}{\texttt{date-8.21}\xspace}
\newcommand{\sedutil}{\texttt{sed-4.1.5}\xspace}
\newcommand{\tarutil}{\texttt{tar-1.14}\xspace}
\newcommand{\nginxutil}{\texttt{nginx}\xspace}

\newcommand{\gziputil}{\texttt{gzip-1.2.4}\xspace}

\newcommand{\etal}{\textit{et al.}\xspace}

\newcommand{\confine}{{\sc Confine}\xspace}

\newcommand{\slimtoolkit}{{\sc SlimToolkit}\xspace}
\newcommand{\speaker}{{\sc Speaker}\xspace}

\newcommand{\paragraphheading}[1]{\noindent\underline{#1}}
\newcounter{issue}
\crefname{issue}{Issue}{Issues}
\Crefname{issue}{Issue}{Issues}

\newif\ifDEBUG
\DEBUGfalse

\ifDEBUG
\newcommand{\SR}[1]{\todo[color=gray,inline]{Sazzadur says: #1}}
\newcommand{\AG}[1]{\todo[color=yellow,inline]{Ashish says: #1}}
\newcommand{\FS}[1]{\todo[color=lime,inline]{Fahad says: #1}}
\newcommand{\MA}[1]{\todo[color=cyan,inline]{Moiz says: #1}}
\newcommand{\MB}[1]{\todo[color=olive,inline]{Bilal says: #1}}
\newcommand{\MK}[1]{\todo[color=pink,inline]{Mohit says: #1}}
\newcommand{\newtext}[1]{\textcolor{red}{#1}}

\else 
\newcommand{\SR}[1]{}
\newcommand{\AG}[1]{}
\newcommand{\FS}[1]{}
\newcommand{\MA}[1]{}
\newcommand{\MB}[1]{}
\newcommand{\MK}[1]{}
\newcommand{\newtext}[1]{\textcolor{black}{#1}}
\fi

\setcopyright{none}

\renewcommand\footnotetextcopyrightpermission[1]{}

\begin{document}

\title{Revisiting Code Debloating with
Ground Truth-based Evaluation}

\lstset{
    basicstyle=\ttfamily\small,
    keywordstyle=\color{black},
    identifierstyle=\color{black},
    commentstyle=\color{black},
    stringstyle=\color{black},
    moredelim=**[is][\color{red}]{@}{@},
    moredelim=**[is][\color{teal}]{@@}{@@},
    moredelim=**[is][\color{magenta}]{@@@}{@@@},
    moredelim=**[is][\color{cyan}]{@@@@}{@@@@}
}

\author{Muhammad Bilal}
\affiliation{\institution{University of Arizona}\country{USA}}
\email{mbilal@arizona.edu}

\author{Moiz Ali}
\affiliation{\institution{University of Illinois Chicago}\country{USA}}
\email{sali269@uic.edu}

\author{Mohit Kumar}
\affiliation{\institution{University of Michigan}\country{USA}}
\email{mohittk@umich.edu}

\author{Fareed Zaffar}
\affiliation{\institution{LUMS}\country{Pakistan}}
\email{fareed.zaffar@lums.edu.pk}

\author{Fahad Shaon}
\affiliation{\institution{Google}\country{USA}}
\email{fahad.shaon@gmail.com}

\author{Ashish Gehani}
\affiliation{\institution{SRI}\country{USA}}
\email{ashish.gehani@sri.com}

\author{Sazzadur Rahaman}
\affiliation{\institution{University of Arizona}\country{USA}}
\email{sazz@arizona.edu}

\begin{abstract}
    Program debloating aims to remove unused code to reduce performance overhead, attack surfaces,  and maintenance costs. Over time, debloating has evolved across multiple layers (container, library, and application), each building on the principles of application-level debloating. Despite its central role, application-level debloating continues to rely on imperfect proxies for measuring performance, such as test-case-driven evaluation for correctness, code size for runtime efficiency, and gadget count reduction for estimating security posture. While there is widespread skepticism about using such imperfect proxies, the community still lacks standardized methodologies or benchmarks to assess the \textit{true} performance of application-level software debloating. This experience paper aims to address the gap.
    
    We revisit the foundations of application-level debloating through a \textit{ground-truth–based} evaluation paradigm. Our analysis of eight state-of-the-art debloaters -- \blade, \chisel, \cov, \cova, \lmcas, \trimmer, \occam, and \razor\ -- uncovers insights previously unattainable through traditional evaluations. These tools collectively span the spectrum of source-to-source, IR-to-IR, and binary-to-binary transformation paradigms, characterizing a holistic reassessment across abstraction levels. Our analysis reveals that while dynamic analysis–based tools often remove up to 94\% of code that should be retained, static analysis–based approaches exhibit the opposite behavior, showing high false retention rates due to coarse-grained dependency over-approximation. Additionally, static analyses may add code by introducing specialized variants of functions. False retentions and removals not only cause functional incorrectness but may also lead to systematic inconsistency, robustness failures, and exploitable vulnerabilities. % Building on these findings, we explore the design space to develop a tool-agnostic mitigation technique that operates on top of existing source-to-source debloaters. We select the source-to-source level because it integrates seamlessly with existing development workflows and supports downstream auditing, testing, and analysis. As a byproduct, we designed \tool, which combines large language models with rule-based mitigations to address the issues we uncovered. Relying on the evaluation of \tool on our ground truth dataset, we discuss the promises and challenges towards making software debloating practical.
\end{abstract}




\maketitle

\pagestyle{plain} % Added for preprint only 

\SR{Double check if the numbers (e.g., number of issues) are consistent ...}

\section{Introduction}

Code bloat occurs when software accumulates features over time, most of which remain unused in typical usage scenarios~\cite{bloat-ref1, bloat-ref2}. Automated software debloating aims to identify and remove unnecessary code (or features) to avoid unnecessary complexity, resource consumption, and potential attack surfaces~\cite{bloat-ref3, bloatref4, occam, bloat-ref7}. Existing debloating paradigms can be organized into three broad categories: application-level (e.g.,~\cite{Chisel, occam, razor, lmcas-arxiv, razor}), library-level (e.g.,~\cite{Piecewise, Saffire}), and container-level (e.g.,~\cite{dockerslim, Cimplifier, ghavamnia2020confine, lei2017speaker}). 
Although the targets differ, existing library- and container-level debloaters are also built on the principles of application-level debloating. This indicates that understanding and improving application-level debloating paradigms in terms of their ability to remove as much \textit{bloat} as possible without compromising functionality or security will directly inform the other paradigms.

Despite playing a central role, the evaluation of existing application-level debloating paradigms largely relies on imperfect proxy metrics. For instance, correctness is typically assessed through test-case–based % functionality testing
evaluation, while performance and security improvements are inferred from reductions in code size or ROP gadget counts.
A critical limitation of such reliance is that these proxy metrics often create a \textit{false sense of accomplishment}. For example, most existing test-case–driven debloaters report substantial reductions in code size and ROP gadget counts, with little to no reported loss in correctness---findings that were later shown to be misleading by others~\cite{debloatbench-sok-gs, brown2024broadcomparativeevaluationsoftware}. This occurs because the test suites used for evaluation frequently fail to capture subtle yet semantically important incorrect behavior. However, test-case-driven debloating paradigms enjoy a widely accepted optimism that providing high-quality and % exhaustive 
extensive test cases can ensure high-quality debloated software. To date, this belief remains unchallenged, as most evaluations in this domain primarily rely on test cases themselves. This dependency on test cases limits the evaluation mechanisms to what the test cases cover.

Similarly, static-analysis–based paradigms~\cite{occam, lmcas-arxiv} often exhibit correct behavior after debloating; however, their corresponding code and ROP gadget reductions tend to be low. In fact, the introduction of multiple specialized variants of functions may increase gadget count. This raises an important open question: are the retained code regions genuinely necessary, or do they merely reflect conservative over-approximations of their underlying analysis engine---a question that remains largely unexplored.

To address this gap, we revisit the foundations of evaluating application-level debloating paradigms and introduce a set of ground-truth–based evaluation metrics. Specifically, we propose a methodology that involves the manual creation of debloated program versions to serve as empirical ground truth. Following this methodology, we created a suite of \napps ground-truth payload applications---which includes widely-studied programs ranging from 5K to 76K LoC. These manually curated versions allow for precise measurement of correctness, performance, and security---providing a more reliable benchmark for assessing the true effectiveness of automated debloating tools. Next, by using the ground-truth–based evaluation, we seek to investigate previously unanswered questions regarding the correctness and efficacy of existing debloating paradigms.
Specifically, by using our ground truth dataset, we comprehensively evaluated $8$ state-of-the-art debloating tools: \blade, \chisel, \cov, \cova, \lmcas, \occam, \razor, and \trimmer. These tools typically operate at two extremes: aggressive \textit{dynamic analysis-based debloaters} that maximize code reduction at the cost of correctness, and conservative \textit{static analysis-based debloaters} that preserve functionality but achieve minimal reduction~\cite{debloatbench-sok-gs, evaldebloatingtools}. \newtext{Since extensive test-case-based evaluation of these tools already exists, we focus our evaluation on the ground truth dataset.}

Our study reveals that despite achieving high code reduction rates, \textit{dynamic analysis-based debloaters} (\blade, \chisel, \cov, \razor) can inadvertently remove up to 94\% of code that should be retained, producing unstable and potentially \textit{insecure} programs, with exploitable vulnerabilities. 

Most importantly, we observed several critical issues that are challenging to capture using test cases. For example, 
essential error logging and handling functions are frequently removed since they are very difficult to exercise with test cases. In multi-threaded programs, we found these debloaters often remove critical thread synchronization primitives, introducing race conditions. These issues may pass the initial test cases but can lead to deadlocks or system crashes in concurrent load environments. 
On the other hand, \textit{static analysis-based approaches} (\lmcas, \trimmer, \occam) prioritize functional correctness of the debloated program, but frequently retain code from 100\% of the functions \textit{not required} in the context of debloating.
\textit{Hybrid} approaches (\cova) attempt to combine dynamic analysis and static analysis, aiming to find the balance between code reduction and functional correctness. Nevertheless, we observed that the limitations of dynamic analysis critically influence the overall performance, as usually static analysis is used around the paths exercised with dynamic analysis, to improve overall coverage.

Our contribution can be summarized as follows:

\begin{itemize}
    \item Development of a comprehensive ground truth dataset for systematic evaluation of source code debloaters, consisting of \napps widely-studied programs ranging from 5K to 76K LoC, including 9 GNU core utilities, 1 \sir benchmark program, and \nginxutil.

    \item Our analysis reveals that while dynamic analysis–based tools often remove up to 94\% of code that should be retained, static analysis–based approaches exhibit the opposite behavior, showing high false retention rates due to coarse-grained dependency over-approximation.
    
    \item Identification of 7 critical issues in existing debloating tools through detailed case studies, out of which only 3 were previously known---underscoring the strength of our ground truth-based evaluation.
%    \item{Introduction of \tool, a tool-agnostic mitigation technique that improves debloated program stability by leveraging both structural analysis-based and LLM powered reasoning-based approaches to address the identified issues.}
\end{itemize}

\section{Ground Truth Creation}
% \MB{GT Debloating is costly. and in theory if debloaters do great on these programs, then they are almost ready for real-world practice. But that is not the case yet.}
\noindent
{\bf Dataset selection.} We created the ground truth dataset by selecting \napps diverse real-world programs. These programs were chosen based on their prior usage in debloating research and the availability of merged all-in-one-file versions (needed for the majority of tested tools) from the authors of previous debloating papers \cite{Blade,Cov,Chisel}. For instance, 9 of these programs are taken from \chiselbench \cite{Chisel},
and 1 each from the SIR benchmark \cite{SIR} and \blade's \cite{Blade} evaluation suite.
Table \ref{tab:benchmark-programs-all} presents the programs we selected for dataset creation and the functionality chosen to create the ground truth. While ground truth debloating is costly, limiting us to \napps programs, our selection strategy prioritizes representativeness over quantity. We choose programs that span different sizes, functionalities, and security contexts. If a debloater struggles on this diverse set, it reveals the fundamental limitations that would manifest across a broader program population.
% Our selection criteria emphasized diversity across multiple dimensions, i.e., size, functional diversity, and security sensitivity of a given program.

\subsection{Debloating Methodology}
\label{subsec:ground-truth-making}
\noindent
{\bf Overview of our manual debloating method.}
Given a deployment context for the specific required functionalities, we start with three primary inputs: (1) the source code file of the program, (2) the list of features to keep, (3) a set of test cases exercising the required functionality of the program. Annotators use these test cases to generate LLVM coverage, helping them better understand the program's execution behavior. The ground truth creation process employs two-phases. The first phase involves main function simplification and understanding of any global state of the program being set in it (\cref{subsubsec:first-phase-ground-truth}). The second phase involves more fine-grained simplification and analysis at the basic block level
(\cref{subsec:second-phase-ground-truth}). Both phases are repeated several times to mitigate any potential oversights and develop a comprehensive understanding of the codebase.

\subsubsection{\textbf{First Phase: Main Function Simplification}}
\label{subsubsec:first-phase-ground-truth}

\noindent
In this phase, we simplify the main function by systematically eliminating unnecessary code paths (based on code for the non-required flags), only preserving paths needed for the required functionality.
The first phase focuses on reducing lines of code in the \emph{main()} function, creating and populating a table for any global flags set, and creating another table to track removed code snippets. %Our primary objective was to simplify the main function by systematically eliminating unnecessary code paths (based on code for the non-required flags), only preserving paths needed for the required functionality. 
By leveraging the input processing logic, we identify and retain paths essential to the required functionality flags. This helps us successfully distinguish the code branches that are relevant to the features to keep or remove. For example, \mkdirutil provides two major functionalities: creating nested directories (using the \texttt{-p} flag) and creating non-nested directories with specific permissions (using the \texttt{-m} flag). In~\cref{fig:mkdir_benchmark_removal_example_1}, we see that through the \emph{getopt\_long()} function, we can identify that the value \emph{112} corresponds to the lower-case \emph{p} character. We identify that the -p flag was not part of the required functionality; therefore, \emph{case\_112} was a non-required code block, and \emph{create\_parents} flag will never be set.

\begin{lstlisting}[
    basicstyle=\ttfamily\scriptsize, 
    linewidth=\linewidth, 
    breaklines=true, 
    breakindent=0pt, 
    caption={LoC that are removed from the ground truth for \mkdirutil when debloating it for the \texttt{-m} flag are highlighted in red. Following the removal, the \emph{make\_path} function becomes unused and is removed.},
    label={fig:mkdir_benchmark_removal_example_1}
]
@@// Parse the input flag to trigger the appropriate path@@
optc = getopt_long(argc, ...));
...
@if (optc == 112) {@ @@// 112 --> `-p' flag @@
    @ goto case_112;
 } @ ...
@@// Corresponding case safe to remove. @@ @
  case_112:  
     create_parents = 1;
     goto switch_break; @
...
if (create_parents) {
@@// Condition should always equal if(0) @@ @
dir = *(argv + optind);
fail = make_path((char const *)dir, ...);@ }
\end{lstlisting}

Next, we perform deadcode elimination to remove all functions that have become unreachable due to the simplification of the main function, record the state of all global variables and flags that were set or unset, and record all the removed code snippets. Moving forward, this helps identify which conditional statements can be simplified to their boolean value based on globally known values, and the purpose of each code snippet, which is kept inside or removed from the ground-truth.

\subsubsection{\textbf{Second Phase: Fine-grained Simplification}}
\label{subsec:second-phase-ground-truth}
In this phase, we comprehensively evaluate whether each line of code directly contributes to the retained functionality, supports critical error-handling mechanisms, or provides optimization strategies for the needed functionality. This phase involves a more detailed code traversal of the remaining functions in a breadth-first manner. Starting from the main function, we trace each of the functions called within it %\newtext{in a breadth-first search manner,} %from top to bottom, 
focusing on understanding the high-level role of each function and the specifics of its implementation. 
For every block (such as an \emph{if-else} or a \emph{while loop} block), sub-blocks, or a line of code (outside of subblocks) inside a function, we make the following set of checks to ensure proper labeling:
\begin{itemize}
    \item Simplify any conditional statements based on knowledge of the global flags we recorded earlier and prune the unreachable branches.
    
    \item Utilize LLVM coverage to trace code paths to understand the calling context of functions under investigation.
    
    \item Investigate how the code paths not in the LLVM coverage relate to the ones in coverage in the context of the required functionality or security. %(See~\cref{sec:app:manualdebloating}).   % refer to figure here
\end{itemize}

We make these decisions based on the understanding of how the conditional statements and corresponding codes relate to either the required functionality or error-handling logic. It is a \textit{time-intensive task} that requires an in-depth understanding of the code base. This process is often supplemented with help from documentation to help understand the code.
% and large language models (LLMs) under circumstances where it becomes challenging to understand the objective of a code snippet. 
% \MB{LLMs were (a) not utilized that much for the ground truth creation process, (b) any output they provided was then either googled for verification from a more reliable source or reasoned about individually, and (c) this creates a weak point for the paper/ground truth methodology creation because it reduces credibility for the annotator} At the end of each function, we note the removed code by appending it to our table of non-required code lines. %In~\cref{sec:app:manualdebloating}, we show an illustrative example of how it is done.
%
Similar to the first phase, we perform deadcode elimination to remove functions that became unreachable after our simplification. The goal is to create a leaner codebase that maintains all the code for the required functionality.

\subsubsection{Validation} 
Before debloating all \napps programs, we conducted a pilot study on \mkdirutil to validate and refine our ground truth creation methodology. During this initial phase, we held weekly meetings with senior authors to review our approach. This iterative process helped us refine our methodology as described in \cref{subsec:ground-truth-making}. Two authors independently created the ground truth for each of the \napps programs, and we validated our refined methodology by measuring their inter-annotator agreement using Cohen's Kappa \cite{cohenkappa}. Cohen's Kappa is a numerical scale representing the level of agreement between two raters, beyond what would be expected by chance. A score of 1.0 is perfect agreement, 0.0 is agreement equivalent to chance, and a score above 0.81 is considered almost perfect. The average Cohen's Kappa score was $\sim$0.93, and the lowest score observed was 0.84 on \chownutil (Details in~\cref{sec:validation}), highlighting the objective nature of the debloating task and the clarity of our manual debloating methodology. % In the next section, we provide a detailed discussion of the debloating phases.

\begin{table}
\footnotesize
\setlength{\tabcolsep}{0.5pt}
\centering
\begin{tabular}{@{}llrr@{}}
\toprule
\textbf{Program} & \textbf{Retained Functionality} & \textbf{LoC\textsubscript{ORG}} & \textbf{LoC\textsubscript{GT} (\%)} \\ \midrule
\mkdirutil & Make dir with permissions (-m) & 5673 & 2967 ({52\%})\\
\uniqutil & Filter adjacent duplicates (no flag) & 8688 & 3151 ({36\%})\\
\bziputil & Compress files (no flag) & 13208 & 6199 ({47\%})\\
\sortutil & Lexicographical sorting (no flag) & 15594 & 5730 ({37\%})\\
\rmutil & Remove files/dirs (-rf) & 7861 & 5729 ({73\%})\\
\dateutil & Time described by string (-d) & 10346 & 7141 ({69\%})\\
\chownutil & Modify ownership (no flag) & 6865 & 3631 ({53\%})\\
\tarutil & Create archive (-cf) & 30824 & 14173 ({46\%})\\
\nginxutil & Serve HTTP static files & 80357 & 34956 ({44\%})\\
\gziputil & Decompress (-d) & 9792 & 4065 ({42\%})\\
\sedutil & Use script file (-f) & 20727 & 19150 ({92\%})\\
\bottomrule
\end{tabular}
\caption{Benchmark programs showing retained functionality, lines of code in the original (LoC\textsubscript{ORG}) and in the ground truth (LoC\textsubscript{GT}), and the percentage of LoC in the ground truth compared to the original. These programs are taken from \chisel~\cite{Chisel}, \blade~\cite{Blade}, and \cov~\cite{Cov}.}
\label{tab:benchmark-programs-all}
\end{table}

\subsection{Manual Debloating Results}

Overall, we achieved substantial LoC reductions in most of the programs we manually debloated. For instance, as shown in~\cref{tab:benchmark-programs-all}, for \nginxutil, we were able to achieve 56\% reduction (45,401 LoC), while for \sortutil, the reduction amounted to 63\% (9,864 LoC). In contrast, the minimal code removal observed for \sedutil, consisting of 8\% (1,577 LoC), occurs due to debloating it on the ``-f" option. This option allows the command script to potentially include any functionality provided by \sedutil, necessitating the retention of most of its code. We refer to these manually debloated programs as \textit{ground truth dataset}.

%\SR{In the following paragraph, tease the downside of using proxy metrices.}
%\underline{At first sight}, these findings highlight an important contrast with existing automated approaches. While certain dynamic-analysis-based debloating tools~\cite{Chisel, razor, Blade} report $\sim$90\% code reduction, such extreme reduction often comes at the cost of correctness and functional integrity. Conversely, static-analysis-based tools~\cite{Chisel, lmcas-arxiv, trimmer} claim that achieving even modest $\sim$20\% reduction demonstrates the effectiveness of their analysis, which in practice reflects overly conservative or imprecise heuristics. Both extremes are concerning, as our manual results indicate that realistic and safe debloating outcomes lie somewhere in between, grounded in semantic understanding rather than blind removal or excessive caution.
% Similarly, for small programs such as \tcasutil, \replaceutil, \scheduleutil, \printtokensutil, and \totinfoutil, we achieved little removal because these programs only contain a singular functionality making most of the program code essential.
% \MB{Maybe the text below is more suited for the discussion section.}It is important to note that for programs with modular code structures, the debloating process tends to be more efficient and achieves significant reductions. For example, in the case of \bziputil, the clear separation between its compression and decompression code allows for the effective removal of redundant code while maintaining functionality. 
\section{Debloating Tool Evaluation}
\label{sec:evaluation-metrics}

In this section, we discuss our quantitative evaluation results for existing application-level debloaters on our ground truth dataset. First, we present our new metrics for quantitative evaluation by using ground truth dataset~\cref{subsection:evaluation:metrics}, followed by the selection process of the tools for evaluation~\cref{subsec:selection-tools}, and, finally the evaluation results~\cref{subsec:ground-truth-comparison}.

\subsection{Evaluation Metrics}\label{subsection:evaluation:metrics}
We evaluate debloating accuracy by comparing debloated programs against the ground-truth dataset at different granularities. Our metrics quantify two error types: \textbf{False Removal (FRm)}, representing code incorrectly removed, and \textbf{False Retention (FRt)}, representing bloat incorrectly retained. The F1-score balances these errors to showcase when a tool leans towards one extreme or the other. Higher F1-scores indicate better overall accuracy, while lower FRm and FRt are individually preferred. We complement these quantitative metrics with qualitative analysis of debloated artifacts in \cref{sec:case_studies}.

\noindent
\textbf{General Evaluation Framework:} We formalize debloating evaluation as a comparison between the debloated program ($P_{d}$) and its ground truth ($P_{gt}$) using a comparator function $\Phi_g$ that operates at a specified granularity $g$.

\begin{equation}
\text{Eval}(P_d, P_{gt}) = \Phi_g(P_d, P_{gt})
\end{equation}

where $g$ defines the unit of comparison (lines of code, functions, IR instructions, etc.). The choice of granularity depends on the debloater category and the feasibility of implementing a meaningful comparison. For granularity $g$, let $U_d^g$ and $U_{gt}^g$ denote code units in $P_d$ and $P_{gt}$, respectively. We define:

\vspace{-12pt}
\begin{align}
\text{FalseRm}^g &= |U_{gt}^g - U_d^g| \\
\text{FalseRt}^g &= |U_d^g - U_{gt}^g|
\end{align}

The corresponding error rates are:
\begin{align}
\text{FRm}_g &= \frac{\text{FalseRm}^g}{|U_{gt}^g|} \times 100 \\
\text{FRt}_g &= \frac{\text{FalseRt}^g}{|U_{base}^g - U_{gt}^g|} \times 100
\end{align}

where $U_{base}^g$ is the original program's unit set. The F1-score balances both errors:

\begin{equation}
\text{F1}_g = \frac{2(100 - \text{FRt}_g)(100 - \text{FRm}_g)}{100(200 - \text{FRt}_g - \text{FRm}_g)}
\end{equation}

By instantiating $\Phi_g$ at different granularities, this framework accommodates various evaluation approaches tailored to different debloater categories, which we discuss next.

\noindent
\textbf{S2S Evaluation.}
For S2S debloaters, the natural unit of measurement is lines of code (LoC). Therefore, we instantiate the general framework with $g = \text{LoC}$. %Line-level granularity enables precise assessment of source-based debloaters by comparing the exact source code units retained or removed against the ground truth.}
% \newtext{For source code-based debloaters, the natural unit of measurement is lines of code (LoC), as these tools operate at source-level granularity. We define the metrics to assess debloating accuracy:}

% \begin{equation}
% \text{FRm}_{LoC} = \frac{\text{\# LoC falsely removed}}{\text{Total LoC retained in ground truth}} \times 100
% \end{equation}

% \begin{equation}
% \text{FRt}_{LoC} = \frac{\text{\# LoC falsely retained}}{\text{Total LoC removed in ground truth}} \times 100
% \end{equation}

% \begin{equation}
% \text{F1}_{{LoC}} = \frac{2(100 - \text{FRt}_{{LoC}})(100 - \text{FRm}_{{LoC}})}{100(200 - \text{FRt}_{{LoC}} - \text{FRm}_{{LoC}})}
% \end{equation}

\noindent
\textbf{IR2IR Evaluation.}
For IR-level debloaters, we employ function-level granularity as the unit of analysis ($g = \text{func}$). This choice is dictated by the fact that IR-level debloaters often rely on compiler-based code specialization passes that distort the original instructions and control-flow structure, making one-to-one correspondence at the instruction-level granularity with the ground truth IR (compiled) infeasible.

\noindent
\textbf{B2B Evaluation.}
For B2B paradigms (such as \razor~\cite{razor}), performing a quantitative evaluation similar to S2S and IR2IR debloaters is challenging. The decompiled binaries are largely distorted from the original code, i.e., function names are stripped and often fragmented---resulting in non-compatible code when compared with the ground-truth dataset, making reliable one-to-one correspondence difficult. Therefore, in this case,  we limit our evaluation to a qualitative assessment of key functional regions.

\begin{table*}[t]
\setlength{\tabcolsep}{5pt}
\centering
\scriptsize
\begin{tabular}{@{}lcccc:cccc:cccc|ccc:ccc|ccc@{}}
\toprule
\textbf{Program} &
\multicolumn{4}{c@{}}{\rotatebox{0}{\textbf{False Removal$_{LoC}$}}} &
\multicolumn{4}{c@{}}{\rotatebox{0}{\textbf{False Retention$_{LoC}$}}} &
\multicolumn{4}{c@{}}{\rotatebox{0}{\textbf{F1 Score$_{LoC}$}}} &
\multicolumn{3}{c@{}}{\rotatebox{0}{\textbf{False Removal$_{func}$}}} &
\multicolumn{3}{c@{}}{\rotatebox{0}{\textbf{False Retention$_{func}$}}} &
\multicolumn{3}{c@{}}{\rotatebox{0}{\textbf{F1 Score$_{func}$}}} \\
\cmidrule(lr){2-5}
\cmidrule(lr){6-9}
\cmidrule(lr){10-13}
\cmidrule(lr){14-16}
\cmidrule(lr){17-19}
\cmidrule(lr){20-22}
 & \rotatebox{90}{\textbf{\blade}}   & \rotatebox{90}{\textbf{\chisel}}
 & \rotatebox{90}{\textbf{\cov}}     & \rotatebox{90}{\textbf{\cova}}
 & \rotatebox{90}{\textbf{\blade}}   & \rotatebox{90}{\textbf{\chisel}}
 & \rotatebox{90}{\textbf{\cov}}     & \rotatebox{90}{\textbf{\cova}}
 & \rotatebox{90}{\textbf{\blade}}   & \rotatebox{90}{\textbf{\chisel}}
 & \rotatebox{90}{\textbf{\cov}}     & \rotatebox{90}{\textbf{\cova}}
 & \rotatebox{90}{\textbf{\lmcas}}   & \rotatebox{90}{\textbf{\trimmer}} & \rotatebox{90}{\textbf{\occam}}
 & \rotatebox{90}{\textbf{\lmcas}}   & \rotatebox{90}{\textbf{\trimmer}} & \rotatebox{90}{\textbf{\occam}}
 & \rotatebox{90}{\textbf{\lmcas}}   & \rotatebox{90}{\textbf{\trimmer}} & \rotatebox{90}{\textbf{\occam}} \\
\midrule
\mkdirutil  & \textbf{90\%} & 67\% & 58\% &  9\% &  4\% &  8\% & 41\% & \textbf{85\%} & 18\% & 49\% & 49\% & 26\%
            &  0\% &  0\% &  0\%
            & \textbf{100\%} & 53\% & 53\%
            &  0\% & 64\% & 64\% \\
\sortutil   & 72\% & 67\% & 56\% & 40\% & 12\% &  7\% & 12\% & 51\% & 42\% & 49\% & 59\% & 54\%
            &  0\% &  0\% &  0\%
            & \textbf{95\%} & 75\% & \textbf{94\%}
            & 10\% & 40\% & 11\% \\
\uniqutil   & 89\% & 78\% & 66\% & 45\% &  3\% & 26\% & \textbf{73\%} & \textbf{79\%} & 20\% & 34\% & 30\% & 30\%
            &  0\% &  5\% &  0\%
            & \textbf{100\%} & 78\% & 83\%
            &  0\% & 36\% & 29\% \\
\rmutil     & 80\% & 68\% & 63\% & 42\% & 22\% & 15\% & 20\% & 63\% & 32\% & 46\% & 51\% & 45\%
            &  0\% &  1\% &  0\%
            & 76\% & 55\% & 55\%
            & 39\% & 62\% & 62\% \\
\bziputil   & 48\% & 66\% & 22\% &  7\% &  5\% &  3\% &  9\% & 17\% & 67\% & 50\% & 84\% & 88\%
            &  0\% &  7\% &  0\%
            & \textbf{100\%} & 87\% & \textbf{100\%}
            &  0\% & 23\% &  0\% \\
\midrule
\chownutil  & \textbf{90\%} & 84\% & 64\% & 34\% &  2\% &  3\% & 12\% & 60\% & 18\% & 27\% & 51\% & 50\%
            &  0\% &  0\% &  0\%
            & 88\% & 77\% & 82\%
            & 21\% & 37\% & 31\% \\
\dateutil   & 45\% & 71\% & 72\% & 35\% & 14\% & 33\% & 29\% & 59\% & 67\% & 40\% & 40\% & 50\%
            & 48\% & 37\% &  0\%
            & 64\% & 72\% & 84\%
            & 43\% & 39\% & 28\% \\
\sedutil    & \textbf{94\%} & 76\% & 58\% & 40\% &  4\% &  5\% & 17\% & 58\% & 11\% & 38\% & 56\% & 49\%
            &  0\% &  0\% &  0\%
            & 29\% & 29\% & 29\%
            & 83\% & 83\% & 83\% \\
\tarutil    & 83\% & \textbf{86\%} & 57\% & 28\% &  8\% &  2\% & 16\% & 45\% & 29\% & 24\% & 57\% & 62\%
            &  0\% &  T  &  0\%
            & \textbf{100\%} & T & \textbf{99\%}
            &  0\% &  T  &  2\% \\
\nginxutil  & 31\% &  T  &  C  &  C  & 23\% &  T  &  C  &  C  & 73\% &  T  &  C  &  C
            &  T  &  T  &  0\%
            &  T  &  T  & \textbf{99\%}
            &  T  &  T  &  2\% \\
\gziputil   & 61\% & 63\% & 33\% &  7\% &  3\% &  3\% &  9\% & 51\% & 56\% & 54\% & 77\% & 64\%
            &  0\% &  T  &  0\%
            & \textbf{100\%} & T & \textbf{98\%}
            &  0\% &  T  &  4\% \\
\midrule
\textbf{Average}
            & 71\% & 73\% & 55\% & 29\% &  9\% & 10\% & 24\% & 57\% & 39\% & 41\% & 55\% & 52\%
            &  5\% &  6\% &  0\%
            & 85\% & 66\% & 80\%
            & 20\% & 48\% & 29\% \\
\bottomrule
\end{tabular}
\footnotesize{\textbf{Legend:} T = Timeout during execution, C = Compilation error in \cov.}
\caption{Comparison of existing debloaters against the ground truth across LoC and function level metrics for our dataset.}
\label{tab:comparison-debloaters}
\end{table*}

\subsection{Selection of Debloating Tools}
\label{subsec:selection-tools}

We select a total of 8 debloating tools---\underline{\chisel}, \underline{\blade}, \underline{\cov}, \underline{\cova}, \underline{\lmcas}, \underline{\trimmer}, \underline{\occam}, \underline{\razor}---for evaluation, based on how they operate across different program representations and employ varying debloating strategies. Next, we discuss our selection process for these tools.

\noindent
\textbf{S2S Tools.} These tools accept source code as input and produce debloated source code as output. Most S2S debloaters (e.g., \chisel~\cite{Chisel}, {\textsc{C-Reduce}\xspace~\cite{Creduce}}, {\textsc{Perses}\xspace~\cite{Perses}}, and \blade~\cite{Blade}) perform \textit{dynamic analysis} using the delta-debugging algorithm for debloating. Another popular \textit{dynamic analysis}-based S2S paradigm leverages code-coverage to avoid overfitting the input test cases (e.g., {\textsc{DomGad}\xspace~\cite{Domgrad}}, {\textsc{DeBop}\xspace~\cite{Debop}}, \cov, and \cova~\cite{Cov}). We select four tools from this category, which represent the state-of-the-art, in terms of code reduction and functional correctness: \blade, \chisel, \cov, and \cova. \blade and \chisel represent delta-debugging–based approaches, whereas \cov represents a coverage-based approach. Building on \cov, \cova represents a \textit{hybrid} approach that leverages static analysis to improve code coverage. % These tools are constrained to operate on a single source code file, and are known for achieving aggressive size reductions.

\noindent
\textbf{IR2IR Tools.}
These tools operate on LLVM bitcode and leverage \textit{static analysis} (i.e., partial evaluation) for code reduction and output reduced LLVM bitcode. Unlike S2S and B2B approaches, IR2IR debloaters do not require test cases. We select \lmcas~\cite{lmcas-arxiv}, \trimmer~\cite{trimmer}, and \occam~\cite{occam} for our analysis as they represent the state-of-the-art for this category. These tools are known for their conservative nature, although \occam supports an optional aggressive specialization by utilizing a hybrid approach with both dynamic and static analysis engines. 

\noindent
\textbf{B2B Tools.}
These tools operate directly on compiled binaries, taking an executable as input and producing a debloated executable as output. B2B tools rely on \textit{dynamic-analysis}-based execution tracing, triggered by user-provided testcases (e.g., \razor~\cite{razor}) or fuzzing (e.g., {\textsc{Ancile}\xspace~\cite{Ancile}}. Due to being open source, we select \razor as our representative tool from this category. \razor runs the binary with user-provided test cases and uses a \textit{Tracer} to collect execution traces. It then decodes the traces to construct the program's CFG, which only contains the executed instructions.

\subsection{Quantitative Evaluation}
\label{subsec:ground-truth-comparison}

\noindent
As discussed in~\cref{subsection:evaluation:metrics}, instantiating a reliable one-to-one correspondence between code debloated with a B2B debloater and ground truth is challenging, which made quantitative evaluation infeasible. Therefore, we limit our quantitative evaluation to S2S and IR2IR tools. 

\noindent
\textbf{Experimental Setup.} To ensure a fair evaluation, we exercised the same functionality across all debloaters. These functionalities were chosen on the basis of the most popular use cases. For test-case–based tools, we gathered relevant tests from their original suites~\cite{Chisel, Cov, Blade} that exercised the functionalities specified in~\cref{tab:benchmark-programs-all} and combined them into a single superset to be used by all tools. \cova uses a configurable parameter, $K$, indicating the depth of the call chain to consider for retention. We tested $10$ values of $K$ values ranging from $1$ to $5$ and $10$ to $50$ with an interval of $10$. We report the scores for $K=50$, which is the biggest value of $K$ reported to be used by the original authors and provides the most optimal results for False Removal (FRm). We report the results of all other tools with the default debloating tool configurations. We present the quantitative evaluation results for all debloaters in~\cref{tab:comparison-debloaters}.

\noindent
\textbf{Results---S2S Tools.}
At the source level, we observe that tools like \blade and \chisel exhibit significantly higher false removal (FRm\textsubscript{$LoC$}) rates compared to coverage-guided debloaters such as \cov and \cova. For instance, \blade removes up to 90\% of essential lines in \mkdirutil and 94\% in \sedutil, suggesting that tools relying solely on dynamic test case execution tend to be overly aggressive in eliminating code not directly linked to observed execution paths. In contrast, \cova achieves the lowest FRm\textsubscript{LoC} values (typically below 45\%), indicating that its coverage-guided dynamic analysis, aided with static analysis, captures a broader range of legitimate functionality. However, this conservatism comes at the cost of higher false retention (FRt\textsubscript{$LoC$}), with up to 85\% in \mkdirutil, demonstrating a key tradeoff of using static analysis for higher functional correctness of the program at the cost of higher bloat.

\noindent
\textbf{Results---IR2IR Tools.}
For IR-level debloaters (\lmcas, \trimmer, and \occam), function-level granularity exposes a markedly different pattern. All three tools maintain near-zero false removals (FRm\textsubscript{func}), suggesting that their conservative static analyses effectively preserve required functionality. However, this safety margin translates into substantial false retention rates, within the 70--100\% range for most programs. 
% In certain cases where we observe lower FRt rates, for example in \mkdirutil and \uniqutil, if we compute a lower bound on false retention at the instruction level—defined as the ratio of instructions in falsely retained functions to the total instructions in all functions that should have been completely removed, we observe substantial retention for both. For \mkdirutil, the ratios are 73.4\% for both \occam and \trimmer, and 100\% for \lmcas; for \uniqutil, the ratios are 94.1\%, 62.2\%, and 99.7\%, respectively. These figures indicate that, even when the number of falsely retained functions is small, they tend to correspond to large, high-impact routines that dominate the remaining code base. 
In cases where the retained function count appears modest, a deeper analysis of instruction-level composition reveals that the falsely retained functions tend to be large and collectively account for the majority of the dead code that should have been eliminated.
These results indicate that IR-level approaches, while robust against functional breakage, tend to over-approximate dependencies at the function boundary and fail to prune finer-grained code related to unrequired functionality.

\begin{tcolorbox}[mysummarybox]
    \textbf{\cref{subsec:ground-truth-comparison}:}
    S2S Tools show upto 94\% FRm$_{LoC}$ (compromising functionality and security), while IR2IR Tools on average  show 77\% FRt$_{func}$ (leaving substantial bloat).
\end{tcolorbox}

% The findings reveal a significant concern: for five of the analyzed programs, all three debloaters exhibit a false removal rate exceeding 50\%. This analysis provides insights to address \textbf{RQ1}: existing source code debloaters often remove excessive amounts of code, leading to a substantial loss of functionality and program integrity. In Section~\ref{sec:case_studies}, we present case studies, looking at the root cause of these high false removal rates by analyzing the methodologies of \blade, \chisel, \cov, and \cova. This analysis helps us identify key limitations in their approaches and create heuristics to address these shortcomings, thereby reducing false removals and improving debloater reliability.
\subsection{Qualitative Evaluation}
\label{sec:case_studies}

\noindent
\textbf{Overview.}
Building upon the quantitative evaluation in \cref{sec:evaluation-metrics} which highlighted the trends in FRm and FRt \textbf{(RQ1)}, this section further presents a qualitative analysis to understand the implications of falsely removing or retaining codes.

\noindent
\textbf{Methodology of Qualitative Analysis.}
\SR{We also studied crashes ...}
This analysis method is divided into two phases: analysis of false removal and analysis of false retention. For each of the phases, we perform a \textit{differential analysis} of every false removal/retention with the ground truth. For each instance of \textit{false removal}, the objective is to pinpoint and categorize whether the removal of this code instance results in adverse outcomes. For each instance of \textit{false retention}, we examine cases where debloating leaves behind partial implementations of functionality that should have been entirely removed---specifically, while the intended functionality is broken, fragments of its supporting logic remain. \newtext{Furthermore, to ground both phases, we ran unseen generality-testing oracles against each debloated binary to observe runtime behavior, using the resulting crashes, hangs, and anomalous outputs as entry points for deeper investigation.}
%The details of the methodology are detailed in \cref{sec:issue-methodology}.

% The original implementations often rely on security checks, input validation, or structural safeguards to ensure correct behavior. If those safeguards are stripped away while fragments of the feature persist, the result is a residual, incomplete code path that can now be exploited.}. 

\noindent
\textbf{Findings Overview.}
Our analysis reveals that functionality and security preservation failures are common across most studied tools, but the nature of the issues varies. Most dynamic analysis-based tools (\chisel, \blade, \cov, \cova, \razor) consistently struggle with maintaining program logic, are subject to residual path vulnerabilities, and produce programs with unpredictable error handling. We also observe tool-specific weaknesses. For instance, \cov and \cova do not include a compilation sanity check in the debloating process and rely on string manipulation only, which can result in syntactic errors in complex code. Similarly, \razor, \cov, \blade, and \chisel's primary reliance on input test cases can lead to improper error logging for system errors that are not invoked during test case execution. Finally, we observe conservative static analysis-based tools (\lmcas, \trimmer, \occam) generally do not suffer from most of the mentioned issues; however, it comes at the tradeoff of extremely high FRt rates. We summarize all the findings about the different issues in~\cref{tab:comparison} and present a detailed analysis below.

\begin{table}[htbp]
    \scriptsize
    \centering
    \begin{threeparttable}
        \caption{Summary of functionality and security issues identified across different tools. Our ground truth based evaluation paradigm surfaced \underline{7 different classes} of functionality and security issues, \underline{out of which only 3 were previously observed} by the testcase based evaluation paradigm.}
        \label{tab:comparison}
        \setlength{\tabcolsep}{1.5pt} % Reduce column spacing
        \small % Reduce font size
        \begin{tabular}{l*{8}{c}}
            \toprule
            \textbf{Observed Issues} & \rotatebox{90}{\chisel} & \rotatebox{90}{\blade} & \rotatebox{90}{\cov} & \rotatebox{90}{\cova} & \rotatebox{90}{\razor} & \rotatebox{90}{\lmcas} & \rotatebox{90}{\trimmer} & \rotatebox{90}{\occam} \\
            \midrule
            \cref{issue:logical-integrity}: Failures in Logical Integrity~\cite{Cov,razor} & $\medbullet$ & $\medbullet$ & $\medbullet$ & $\medbullet$ & $\medbullet$ & - & - & - \\
            \cref{issue:residual-paths}: Residual Path Vulnerabilities~\cite{razor} & $\medbullet$ & $\medbullet$ & $\medbullet$ & $\medbullet$ & $\medbullet$ & - & - & - \\
            \cref{issue:deletion-sensitive-files}: Unsafe Intermediate State Execution & - & $\medbullet$ & - & - & - & - & - & - \\
            \cref{issue:Thread-Synchronization}: Thread Synchronization & $\medbullet$ & $\medbullet$ & - & - & - & - & - & - \\
            \cref{issue:Unpredictable-error-handling}: Unpredictable Error Handling~\cite{razor} & $\medbullet$ & $\medbullet$ & $\medbullet$ & $\medbullet$ & $\medbullet$ & - & - & - \\
            % \cref{issue:error-logging-system-errors}: Error-Logging for System Errors & $\medbullet$ & $\medbullet$ & $\medbullet$ & $\medbullet$ & $\medbullet$ & - & - & - \\
            \cref{issue:7}: Variable State Management & $\medbullet$ & $\medbullet$ & $\medbullet$ & $\medbullet$ & - & - & - & - \\
            \cref{issue:Debloated-program-not-compiling}: Debloated program not compiling & - & - & $\medbullet$ & $\medbullet$ & - & - & - & - \\
            \bottomrule
        \end{tabular}
    \end{threeparttable}
\end{table}

\refstepcounter{issue}\label{issue:logical-integrity}
\subsubsection*{Issue 1: Failures in Logical Integrity}

\noindent
\blade, \chisel, \cov, and \razor exhibit critical failures in preserving logical integrity under following three categories:

\noindent
\paragraphheading{Merging mutually exclusive blocks.} \blade merges code from mutually exclusive \textit{if-else} branches, forcing execution of logic never intended to run together. \cref{merging_exclusive_blocks} (Appendix~\ref{code-appendix}) demonstrates this in \rmutil, where red and blue blocks originally separated by control flow are merged into a single path, compromising program integrity.

\noindent
\paragraphheading{Forcing independent blocks into nested structures.} \razor transforms independent conditional blocks into nested dependencies by selectively removing code. Originally separate \textit{if} conditions become contingent on preceding conditions, creating unintended execution paths. \cref{fig:razor-sort-if-nest} shows this in \sortutil, where independent checks for \texttt{1UL < nhi} and \texttt{1UL < nlo} are forced into a nested structure under a single \textit{else} statement.

% \begin{figure}
%     \centering
%     \begin{subfigure}[t]{0.45\textwidth}
%         \centering
%         \begin{lstlisting}[basicstyle=\ttfamily\scriptsize, linewidth=\linewidth, breaklines=true, breakindent=0pt, numbers=left,  numberstyle=\scriptsize]
% if (nthreads > 1UL) { ... } else {
%     nlo = node->nlo;
%     nhi = node->nhi;
%     temp = (struct line *)(lines - total_lines);
%     if (1UL < nhi) { 
%       sequential_sort(lines - nlo, ...);
%     @}@
%     if (1UL < nlo) {
%       sequential_sort(lines, ...);
%     @}@
%     node->lo = (struct line *)lines;
%     node->hi = (struct line *)(lines - nlo);
%     node->end_lo = (struct line *)(lines - nlo);
%     node->end_hi = (struct line *)((lines - nlo) - nhi);
%     queue_insert(queue, node);
%     merge_loop(queue, total_lines, tfp, temp_output);
% @}@
% pthread_mutex_destroy(& node->lock);
% return;
%         \end{lstlisting}
%     \end{subfigure}
%     \caption{\newtext{\razor transforms independently executing \textit{if} blocks into nested structures by selectively removing code, fundamentally altering control flow. Example from \sortutil's \textit{sortlines} function shows removed lines in red.}}
%     \label{fig:razor-sort-if-nest}
% \end{figure}

\begin{lstlisting}[
    basicstyle=\ttfamily\scriptsize, 
    linewidth=\linewidth, 
    breaklines=true, 
    breakindent=0pt, 
    numbers=left,  
    numberstyle=\scriptsize,
    xleftmargin=2.3em,
    caption={Example from \sortutil's \textit{sortlines} function shows \razor transforms independently executing \textit{if} blocks into nested structures by selectively removing the closing brackets (highlighted in red). This matches the opening bracket on line 2 and closing bracket on line 16, fundamentally altering control flow from its original intention.},
    label={fig:razor-sort-if-nest}
]
...    
    if (nthreads > 1UL) { ... } else {
        nlo = node->nlo;
        ...
        if (1UL < nhi) { 
          sequential_sort(lines - nlo, ...);
        @}@
        if (1UL < nlo) {
          sequential_sort(lines, ...);
        @}@
        ...
        merge_loop(queue, total_lines, ...);
    @}@
    pthread_mutex_destroy(& node->lock);
    return;
}
\end{lstlisting}

\noindent
\paragraphheading{Breaking basic block integrity.} The approach of removing code with the granularity of a single line overlooks the logical dependencies within a block (as in \cref{removing_single_line} in \cref{code-appendix}). Code blocks or compound statements represent a single unit of logic. Splitting these blocks further from within can potentially cause stability issues and erroneous behavior. This phenomenon can also result in eliminating only the \textit{else} statements of \textit{if-else} blocks and only the \textit{while} loop statements, or break conditions from the loop structure (\cref{fig:removing_loop_break} in~\cref{code-appendix})).

\noindent
\paragraphheading{Removing critical dependencies.} These debloaters lack the capacity to reason about inter-component dependencies, leading to the removal of essential validation checks. \chisel removes security checks from \rmutil's \emph{fts\_read()} function (\cref{fig:rm-fts-read-securitycheck} in \cref{code-appendix}), eliminating file tree entry existence verification and error state validation. This allows the debloated program to proceed with file deletions even when the file tree state is invalid, introducing serious vulnerabilities.

\refstepcounter{issue}\label{issue:residual-paths}
\subsubsection*{Issue 2: Residual Path Vulnerabilities}

Invoking debloated programs with unsupported flags or inputs can activate residual code paths containing partially removed or unintentionally retained code, leading to crashes, infinite loops, or security vulnerabilities. Although conservative dynamic analysis S2S debloaters (\cov, \cova) retain extra code to preserve functionality, they may inadvertently keep dangerous paths that enlarge the attack surface.
Specifically, \cova introduces heuristics atop \cov to improve correctness, but these heuristics often retain broken and therefore potentially harmful paths. Conservative thresholds (e.g., $k=30$) also include code from unused functionality with broken dependencies. As shown in \cref{fig:gzip-conservative-covA}, \emph{gzip-1.2.4} debloated for decompression still retains compression code, and invoking it with compression flags triggers incomplete paths that delete input files without producing output. Similarly, \cref{fig:date-conservative-covA} (\cref{code-appendix}) shows that the \emph{date} utility debloated to display system time incorrectly retains \emph{--reference} functionality, allowing unauthorized access to file modification times.
\begin{lstlisting}[
    basicstyle=\ttfamily\scriptsize, 
    linewidth=\linewidth, 
    breaklines=true, 
    breakindent=0pt, 
    caption={Retained compression code in \emph{gzip-1.2.4} debloated for decompression using \cova (k=30). Invalid inputs trigger this path, causing \underline{original file deletion} without creating compressed output.},
    label={fig:gzip-conservative-covA}
]
chown(ofname, ...);
remove_ofname = 0;
chmod(ifname, ...);
tmp___0 = unlink(ifname); @@//delete the file@@                        
\end{lstlisting}

\refstepcounter{issue}\label{issue:deletion-sensitive-files}
\subsubsection*{Issue 3: \newtext{Unsafe Intermediate State During Debloating}}

\newtext{The test-oracle paradigm assumes that (1) test execution on intermediate states is safe, and (2) a passing test suite is sufficient evidence that a removal is semantically benign. Both assumptions break down for safety-critical guard logic.}
\newtext{\blade removes code from \rmutil's \emph{fts\_build} function (in the \texttt{gnulib} library) that prevents infinite recursion by skipping \texttt{`.'} and \texttt{`..'} entries during traversal. The consequence manifests during the debloating process itself: when \blade removes this guard and executes the test suite on the intermediate state, the broken traversal logic deletes the container's \texttt{/bin} directory, crashing the process before completion (\cref{rm_bin_deletion_example1} in \cref{code-appendix}).}
\newtext{Such guard logic has no observable effect under normal conditions as its failure modes are environment-dependent which are unlikely to be covered by any practical test suite. The paradigm's equation of "test passes" with "removal is safe" is therefore fundamentally unsound for guard logic: a removal can silently pass all tests while introducing catastrophic behavior, or as here, destabilize the tool's own execution environment.}

\MB{Renamed the issue, and removed information the following information about trimmer: \trimmer removes the \emph{get\_root\_dev\_ino} function from \rmutil, which is responsible for obtaining the device and inode fingerprint of the root directory. Improper handling of this routine could lead to unsafe file operations, such as the accidental deletion of the root filesystem. Empirically, the debloated \trimmer binary crashes on all tested inputs, confirming that its output is not functionally correct.}
\SR{We can add this issue to Issue 1? (Although, Issue 1 already seems big)}

\refstepcounter{issue}\label{issue:Thread-Synchronization}
\subsubsection*{Issue 4: Thread Synchronization}
% \label{subsection:Thread-Synchronization}

Debloaters like \chisel and \blade assume code is safe to remove if it does not trigger runtime issues during tested execution paths. This premise is particularly dangerous for multi-threaded programs, where these tools inadvertently remove critical synchronization primitives such as mutex locks, condition variables, and coordination flags. Such removals may not manifest during testing, as race conditions and deadlocks often appear only under specific timing conditions or heavy load. This limitation highlights these tools' inability to recognize the semantic importance of a particular coding construct.

\cref{fig:sort-thread-unsafe} demonstrates this in \sortutil, where \blade removes thread synchronization from \emph{queue\_insert()}, a function where multiple threads concurrently insert merge tasks into a shared priority queue. The removed mutex operations leave the queue susceptible to concurrent access corruption, while eliminating the queued flag and condition variable signal breaks worker thread coordination, causing missed work or deadlocks.

\begin{lstlisting}[
    basicstyle=\ttfamily\scriptsize, 
    linewidth=\linewidth, 
    breaklines=true, 
    breakindent=0pt, 
    numberstyle=\scriptsize,
    caption={{Thread synchronization code (highlighted in red) removed by \blade from \sortutil's merge sort implementation, eliminating thread-safe logic.}},
    label={fig:sort-thread-unsafe}
]
static void queue_insert(...){
    @pthread_mutex_lock(&queue->mutex);@
    heap_insert(queue->priority_queue, ...);
    @node->queued = (_Bool)1;
    pthread_mutex_unlock(&queue->mutex);
    pthread_cond_signal(&queue->cond);@ 
}
\end{lstlisting}

\refstepcounter{issue}\label{issue:Unpredictable-error-handling}
\subsubsection*{Issue 5: Removal of Error-Handling Code}
Dynamic analysis-based debloaters systematically remove error-handling infrastructure because such code is rarely executed during testing. This manifests across two classes of error paths -- Environment-dependent and Input-triggered cases. Both of these cases share the same root cause: the test oracle can only observe what the test environment exposes.

\paragraphheading{Environment-dependent error handlers.} This class of error paths cannot be triggered through user input at all, as their activation depends on environmental state rather than program behavior.~\cref{fig:quote_call_mkdir} illustrates this in \mkdirutil, where \emph{quote()} is invoked when the \emph{mkdir} system call fails. Triggering this path requires inducing a kernel-level failure (e.g., exhausting inodes, simulating disk faults), which no existing debloating test suite attempts. % We observe a similar phenomenon in ~\cref{fig:rm-unpredictable-error}, which shows \rmutil where handlers managing \emph{unlinkat} failure scenarios are removed by \chisel, \blade, and \razor. 
This points to a fundamental mismatch between what dynamic observability can capture and what must be preserved for robustness. Existing test suites keep the execution environment fixed. They exercise program functionality across anticipated and unanticipated user inputs, but never manipulate the environment itself to observe how the program responds to system call failures. As a result, any code whose activation depends on the environment is invisible to the debloater's oracle. Therefore, this is not a coverage problem that is solvable by writing more tests; it requires a conceptual shift from testing program behaviors in isolation.

\begin{lstlisting}[
    basicstyle=\ttfamily\scriptsize, 
    linewidth=\linewidth, 
    breaklines=true, 
    breakindent=0pt,
    numberstyle=\scriptsize,
    caption={{Error-logging code (highlighted in red) removed by \cov from \mkdirutil. The \emph{quote} function on line 8 is triggered only when \emph{mkdir} system call fails, but gets removed due to lack of test coverage.}},
    label={fig:quote_call_mkdir}
]
dir___0 = (char const *)*(...);
fail = make_dir(dir___0, ...); @@//wrapper for mkdir@@ @
if (!fail) { @@ // If the system call fails @@
 if (!create_parents) {
  if (!dir_created) {@ @@ // log errors and abort. @@ @
    tmp___6 = quote(dir___0);
    tmp___7 = gettext("cannot create directory %s");
    error(0, 17, ...) }}} @
\end{lstlisting}

% \begin{lstlisting}[
%     basicstyle=\ttfamily\scriptsize, 
%     linewidth=\linewidth, 
%     breaklines=true, 
%     breakindent=0pt, 
%     numberstyle=\scriptsize,
%     caption={{Critical error handling for \texttt{unlinkat} system call (highlighted in red) removed by \chisel, \blade, and \razor from \rmutil due to infrequent execution during testing.}},
%     label={fig:rm-unpredictable-error}
% ]
% tmp___4 = unlinkat(fts->fts_cwd_fd, ...);
% if (tmp___4 == 0) { ...
%   return ((enum RM_status)2);
% }@
% tmp___8 = __errno_location();
% if (*tmp___8 == 30) { ... } @
% ...  @@// Series of more error checks @@ @
% \end{lstlisting}

\paragraphheading{Input-triggered error handlers.} Error handlers reachable through user input are removed when test suites fail to exercise the corresponding failure conditions. While better test suites could, in principle, cover some of these paths, doing so requires explicitly designing tests around failure conditions. The challenge of creating such test cases is often more challenging than writing the code itself, which is often underestimated in the debloating context. While our analysis shows that these error handlers are usually responsible for error reporting, their presence is critical to ensure that the program does not enter an invalid program state, as it was originally intended to. For example, \cref{fig:usage-function-exit-agressive} shows \uniqutil debloated for basic duplicate removal, where the original \emph{usage(1)} call on invalid input (which displays usage information and terminates with \emph{exit(1)}) is removed. This causes the program to continue into unintended code paths, resulting in an infinite loop and resource exhaustion, but may expose additional attack surfaces depending on the code paths entered. While \razor at least aborts when entering pruned code, \chisel and \blade inadvertently remove error-handling logic that prevents execution continuation under error conditions.

\begin{lstlisting}[
    basicstyle=\ttfamily\scriptsize, 
    linewidth=\linewidth, 
    breaklines=true, 
    breakindent=0pt, 
    numberstyle=\scriptsize,
    caption={{Error-handling logic from \uniqutil's \emph{main} function, removed by \blade, \chisel, and \razor during aggressive debloating.}},
    label={fig:usage-function-exit-agressive}
]
if (nfiles == 2) {
  tmp___4 = quote((char const   *)optarg);
  tmp___5 = gettext("extra operand %s");
  error(0, 0, (char const   *)tmp___5, tmp___4);
  usage(1);
}
\end{lstlisting}
 
% As a consequence of the narrow focus on functionality of the program, most of the current approaches do not value the importance of error-logging. The \emph{quote()} family of functions, which is used for error handling across most programs in our benchmark, accounts for $\sim$1200 LoC and is completely removed by all of the debloaters of this category (\chisel, \blade, \cov, and \cova).

This reveals a significant methodological flaw in current debloating strategies. Even with comprehensive test suites, there are essential program components (particularly error-handling routines) that may not be exercised during typical usage scenarios but remain crucial for robustness.

\refstepcounter{issue}\label{issue:7}
\subsubsection*{Issue 6: Variable State Management}
Debloating tools can create serious security vulnerabilities by removing critical variable initialization code, leading to undefined behavior through two patterns: propagating uninitialized values and eliminating state initialization.

\noindent
\paragraphheading{Direct propagation of uninitialized values.} \blade removes the sole assignment to variable \emph{tmp\_\_\_2} in \dateutil (\cref{fig:uninitialized_variable} in Appendix~\ref{code-appendix}), causing the function to return arbitrary memory content. This undefined behavior enables unexpected program states.

\noindent
\paragraphheading{Critical state initialization removal.} \cref{fig:uninitialized_variable_rm} shows \chisel removing \emph{rm\_option\_init} and subsequent state initializations from \rmutil debloated with \texttt{-rf} flags. The \emph{rm\_options} structure contains security flags and settings for interactive mode. Without proper initialization, the program operates with undefined security states, enabling unsafe file operations or privilege escalation.

\begin{lstlisting}[
    basicstyle=\ttfamily\scriptsize, 
    linewidth=\linewidth, 
    breaklines=true, 
    breakindent=0pt, 
    numberstyle=\scriptsize,
    caption={{Remnants of \rmutil's main function debloated by \chisel for \texttt{-rf} flags. Five of 214 removed lines (highlighted in red) handle critical state initialization for secure file removal.}},
    label={fig:uninitialized_variable_rm}
    ]
int main(int argc, char **argv) {
  struct rm_options x;
  char **file;
  enum RM_status tmp___8;@
  rm_option_init(&x);
  x.interactive = (enum rm_interactive)5;
  x.ignore_missing_files = (_Bool)1;
  prompt_once = (_Bool)0;
  x.recursive = (_Bool)1;@
  file = argv + optind;
  tmp___8 = rm((char *const *)file,...); }
\end{lstlisting}

\noindent
\paragraphheading{Memory corruption through uninitialized pointers.} \blade's debloating of \mkdirutil with \texttt{-m} flag removes variable initializations, passing an uninitialized pointer to the \emph{strtoul} system call (\cref{fig:mkdir-memory-leak} in \cref{code-appendix}). This corrupts \emph{argv[3]}, removing its null terminator and causing directory creation with arbitrary names that expose program memory.

While the issue of variable state management is more prevalent in more aggressive debloaters, it is observed in \cov and \cova as well, highlighting a fundamental limitation of current debloating approaches: the inability to recognize the security implications of removing initialization code. Such removals may not trigger immediate crashes or test failures, yet create latent vulnerabilities exploitable in production environments. 

\refstepcounter{issue}\label{issue:Debloated-program-not-compiling}
\subsubsection*{Issue 7: Complex Control Structures}

\cov and \cova fail to maintain syntactic correctness when processing complex control structures, particularly in programs outside their primary evaluation suite. Debloating \nginxutil produces 1,489 compiler errors, revealing fundamental flaws in their code removal approach.
\cov removes entire \textit{case} blocks while retaining labels, creating syntactically invalid C code with orphaned \textit{case} statements. Additionally, \cov fails to maintain bracket matching when removing complex blocks, further compromising syntactic validity.

\textit{This reflects a broader methodological issue:} \cov relies on predefined rules, pattern matching, and regular expressions for code removal without compilation validation. Unlike tools incorporating compilation checks during debloating, \cov compiles only at initialization to exercise test cases and build the AST, then applies hard-coded rules without verifying syntactic validity. This approach fails for programs with edge cases in control structure handling, rendering debloated output to be non-compileable.
\begin{tcolorbox}[mysummarybox]
% \footnotesize
\textbf{\cref{sec:case_studies}:}
Dynamic analysis-based tools frequently break control-flow logic, disrupt state dependencies, corrupt synchronization patterns, and remove critical error-handling or edge-case behaviors.  Static analysis-based tools avoid these failures.

\end{tcolorbox}
\subsection{Implications of our Findings}
\label{subsec:insights-existing-paradigms}

Overall, our results demonstrate a clear tension between soundness and completeness in debloating strategies. While dynamic source code level techniques (\blade, \chisel, \cov) often achieve significant reductions in code size, the resulting debloated programs suffer from severe correctness issues. They risk over-pruning due to incomplete semantic understanding, leading to substaintial false removals when evaluated against ground truths.

Static analysis-based IR-level tools (\lmcas, \trimmer, \occam) sit at the opposite extreme with highly conservative behavior. Their broad over-approximation ensures correctness is preserved, but also causes them to retain large portions of code that are not required for the targeted functionality. This leads to substantial false retention and limits the overall reduction that they can achieve.

Hybrid approaches that combine static and dynamic analysis try to overcome the weaknesses of both techniques, but they bring their own issues. For example, \cova builds on \cov by using static analysis heuristics to identify and keep code similar to what the dynamic analysis preserved. However, since these heuristics depend on the dynamic analysis results, any gaps or biases in the runtime coverage end up being carried over to the statically retained code, effectively locking the same limitations into the final debloated binary.

\section{Related Work}

\noindent
\textbf{Alternate domains of debloating.}
Our findings have implications for debloating in other domains beyond C/C++ applications.
For example, Azad~\etal~\cite{bloat-ref3} debloated PHP web applications by instrumenting PHP files and observing their usage with a large set of user requests. \confine~\cite{ghavamnia2020confine}, \speaker~\cite{lei2017speaker}, and \slimtoolkit~\cite{dockerslim} debloat docker containers by observing system call execution. While Hassan \etal~\cite{container-debloaters} showed major deficiencies in the correctness of these tools, their evaluation also relied on test cases, making it subject to the same limitations we identified for application-level debloating.

\noindent
\textbf{Systematization of debloating knowledge.} Several studies have examined the capabilities of the debloating landscape. Brown \textit{et al.} conducted the largest comparative evaluation to date, assessing ten tools across twenty benchmarks and twelve metrics, and found that only 22\% of debloating attempts succeeded on high-complexity benchmarks, with 13\% yielding sound outputs \cite{evaldebloatingtools}. Ali \textit{et al.}'s DebloatBench$_A$ integrated four representative tools (\chisel, \occam, \razor, and {\textsc{Piece-wise}\xspace}) across multiple abstraction levels, revealing that static tools ensured correctness while dynamic ones struggled, especially as program complexity increased \cite{debloatbench-sok-gs}. Alhanahnah \textit{et al.} proposed a multilevel taxonomy of debloating workflows and identified open challenges such as robustness, SBOM integration, machine learning assistance, and CI/CD compatibility~\cite{DBLP:conf/feast/AlhanahnahBG24}. In contrast to these test-case-driven studies, our work proposes a ground-truth-based evaluation, allowing for a direct quantitative and qualitative analysis of the performance. Specifically, our analysis revealed $7$ new classes of debloating issues, out of which only $3$ were previously known. 

\section{Discussion}

\noindent

\noindent
{\bf Implications of our work.} This work establishes a new foundation for evaluating debloaters by introducing a ground-truth dataset containing \napps diverse real-world programs. Building on this benchmark, this study shows critical deficiencies of existing dynamic-analysis-based source code debloaters (\cref{sec:case_studies}). We show that even with carefully constructed test case suites, capturing all edge cases and required code paths (i.e., thread synchronization or kernel error handling codes) is infeasible. This limitation resulted in critical functional incorrectness, removal of error-handling codes, and even incidental vulnerabilities in the debloated code. % While our evaluation of \tool shows early promises of how LLMs can help, it also shows room for improvement -- specifically, in reducing false retention cases.

\noindent
{\bf Threats to Validity.}
During our ground truth dataset construction, we thoroughly validated and calculated inter-annotator agreement which showed a minimal impact of subjectivity bias (\cref{tab:ground-truth-evaluation}); however, human error could inadvertently introduce biases.
% Moreover, our tool relies on the accuracy of the underlying LLMs. \tool's performance might suffer in handling program semantics that the LLM has not encountered during its training phase. 
% Additionally, even with state-of-the-art LLMs like \geminipro (2M token context), \tool can only process programs up to $\sim$150K lines of code~\cite{gemini-contextwindow}. While increasing the context window can theoretically enable the processing of larger codebases, it may also degrade performance due to the increased computational overhead and difficulty in maintaining focus across extensive inputs. This indicates a need for future efforts from the debloating community
% \SR{We just criticized Leader on this}.
% \newtext{Finally, while LLMs with large context windows (e.g., \geminipro with its 2M token context) can in principle leverage more context as an input for reasoning, we observed that increasing input size does not monotonically improve performance. Longer contexts introduce computational overhead and dilute the model's focus, ultimately hurting accuracy. Addressing this tension between context richness and inference quality remains an open challenge for the debloating community.}

\noindent
\textbf{Future Direction.} In principle, it may be feasible to use our ground truth dataset to evaluate LLVM IR- or binary-level debloaters with fine granularity. However, unambiguously mapping instructions from the IR or binary code to the corresponding source location poses a nontrivial challenge, which deserves independent treatment. 
% As a proof of concept, we conducted preliminary evaluations using \occam \cite{occam} (IR level) and \razor \cite{razor}  (binary level), facilitated through reverse engineering of binaries. A comprehensive treatment of such low-level debloaters remains outside the scope of this paper.
% Thus, we consider the problem of evaluating IR- or binary-level debloaters as outside the scope of this paper.

\section{Conclusion}
In this study, we created a new ground-truth-based paradigm for evaluating the performance of application-level debloaters. We created a manually debloated ground truth dataset of \napps real-world programs and extensively evaluated the security and reliability aspects of programs debloated with $8$ state-of-the-art source code debloaters.
Our evaluation highlights a significant limitation in test case-based evaluation of program debloaters, as certain critical aspects of code are infeasible to capture with test cases. 

\appendix

\section{Ground truth validation}\label{sec:validation}

\begin{table}[!h]
\footnotesize
\centering
\begin{tabular}{|l|c|}
\hline
\textbf{Program} & \textbf{Agreement Cohens' Kappa} \\ \hline
\mkdirutil & 0.980  \\ \hline
\sortutil & 0.975  \\ \hline
\uniqutil  & 0.942  \\ \hline
\rmutil & 0.905  \\ \hline
\bziputil & 0.937  \\ \hline
\chownutil & 0.840 \\ \hline
\dateutil & 0.920 \\ \hline
\sedutil & 0.912 \\ \hline
\tarutil & 0.916 \\ \hline
\nginxutil & 0.938 \\ \hline
\gziputil & 0.940 \\ \hline
\end{tabular}

\caption{Inter-annotator agreement of ground truth creation of \napps programs debloated manually.}
\label{tab:ground-truth-evaluation}
\end{table}

Our ground truth benchmark achieves high Cohen's kappa scores (\cref{tab:ground-truth-evaluation}), demonstrating that program debloating is largely objective when given specific requirements. \cref{fig:rm_benchmark_removal_example} illustrates a typical case where both annotators easily identify and agree on removable code. This consistency stems from our structured approach: annotators first simplify the main function and establish which flags must remain set throughout the program. These predetermined flags then guide the systematic identification of removable code snippets.

\begin{lstlisting}[
    basicstyle=\ttfamily\footnotesize, 
    linewidth=\linewidth, 
    breaklines=true, 
    breakindent=0pt, 
    numberstyle=\footnotesize,
    caption={LoC that are removed from the ground truth for \rmutil when debloating it for the \emph{-rf} flag are highlighted in red.},
    label={fig:rm_benchmark_removal_example}
]
static enum RM_status excise(FTS *fts, ...) {
...
@@ // path only traversed if verbose flag was set in the main function @@  @
if (x->verbose) {
  tmp___0 = quote((char const *)ent->fts_path);
  if (is_dir) {
    tmp___1 = gettext("removed ...);
    tmp___3 = tmp___1;
  } else {
    tmp___2 = gettext("removed %s\n");
    tmp___3 = tmp___2;
  }
  printf((char const *)tmp___3, ...);
} @
\end{lstlisting}

\cref{fig:chown-validation-2} demonstrates the primary sources of inter-annotator disagreement encountered across our ground truth programs. The first example involves an unused struct \texttt{LCO\_ent}, which one annotator overlooked during analysis. Such oversights are possible given that programs often contain multiple structs that conventional deadcode removal tools fail to identify. The second example shows repeated extern header statements in the merged program, another category that deadcode removal tools frequently miss. Although annotators perform multiple manual iterations over each program, the substantial size of our benchmark programs can make these specific cases challenging to consistently identify, leading to occasional disagreements between annotators.

\begin{lstlisting}[
    basicstyle=\ttfamily\footnotesize, 
    linewidth=\linewidth, 
    breaklines=true, 
    breakindent=0pt, 
    numberstyle=\footnotesize,
    caption={LoC removed by only one annotator in \chownutil's ground truth creation. Most frequent disagreements across programs between annotators were observed.},
    label={fig:chown-validation-2}
]
@@// Unused struct, following removal of other code @@
struct LCO_ent {
    dev_t st_dev;
    _Bool opt_ok;
};
...@@
// Merged header that is not being used anymore@@
extern __attribute__((__nothrow__)) int(__attribute__((__leaf__)) fchdir)(int __fd);
\end{lstlisting}

% Our ground-truth creation process employed a two-pass, manual debloating methodology. Given an oracle, we started with three primary inputs: the original merged code file, training cases for debloating, and the program's LLVM coverage providing line-level insights into its execution.

For another example of our approach, consider debloating process for \rmutil with the \emph{-rf} flag. During the first pass, paths to other flags, such as the verbose flag, were removed from the main function, and we recorded the global state of \emph{x--$>$verbose} as 0. However, several other functions continue to contain conditions that check for such flags. \cref{fig:rm_benchmark_removal_example} shows how a condition checking for the verbose condition was removed in the \texttt{exise} function during the second pass.
% We used two key resources during this process. LLVM coverage data gave us insights into actual code execution during test cases. Additionally, we consulted documentation to understand the roles of library functions, ensuring we didn't accidentally remove critical logic.

\begin{figure}
    \centering
    % Original and Debloated subfigures side by side
    \begin{subfigure}[t]{0.45\textwidth}
        \centering
        \begin{lstlisting}[basicstyle=\ttfamily\footnotesize, linewidth=\linewidth, breaklines=true, breakindent=0pt, numbers=left,  numberstyle=\scriptsize]
if (!(sp->fts_options & 4)) {
@ if (sp->fts_options & 512) {
    if (sp->fts_options & 512) {
      tmp___20 = -100;
    } else {
      tmp___20 = sp->fts_rfd;
    }
    cwd_advance_fd(...);
    tmp___23 = 0;
  } @ @@@@
  else {
    ...
    tmp___22 = fchdir(tmp___21);
    tmp___23 = tmp___22;
  }
  ...
} @@@@
        \end{lstlisting}
        \caption{Original Code}
    \end{subfigure}
    \hfill
    \begin{subfigure}[t]{0.45\textwidth}
        \centering
        \begin{lstlisting}[basicstyle=\ttfamily\footnotesize, linewidth=\linewidth, breaklines=true, breakindent=0pt, numbers=left,  numberstyle=\scriptsize]
if (!(sp->fts_options & 4)) { 
    @tmp___20 = -100;
    cwd_advance_fd(...);
    tmp___23 = 0; @
@@//else statement removed and the mutually exclusive blocks are merged@@
    @@@@tmp___22 = fchdir(tmp___21);
    tmp___23 = tmp___22; @@@@
}
        \end{lstlisting}
        \caption{Debloated Code}
    \end{subfigure}

    \caption{Mutually exclusive LoC are merged into one execution path for \rmutil with \blade. Red and blue represent mutually exclusive blocks of code that are merged into the same execution path in the debloated file.\SR{Candidate for removal.}}
    \label{merging_exclusive_blocks}
\end{figure}

\section{Supplementary Code Examples of Observed Issues}
\label{code-appendix}

\begin{lstlisting}[
    basicstyle=\ttfamily\footnotesize, 
    linewidth=\linewidth, 
    breaklines=true, 
    numberstyle=\footnotesize,
    caption={\blade removes individual lines from basic blocks in \texttt{\dateutil}, breaking logical dependencies within code units (highlighted in red).},
    label={removing_single_line}
    ]
if (!pc.days_seen) {
  tm.tm_hour = to_hour(pc.hour, ...);
  if (tm.tm_hour < 0) { goto fail; }
@  tm.tm_min = (int)pc.minutes;  @
  tm.tm_sec = (int)pc.seconds.tv_sec;
} else {
  tm.tm_sec = 0;
  tm.tm_min = tm.tm_sec;
@  tm.tm_hour = tm.tm_min; @
  pc.seconds.tv_nsec = (__syscall_slong_t)0;
}
\end{lstlisting}

\begin{lstlisting}[
    basicstyle=\ttfamily\footnotesize, 
    linewidth=\linewidth, 
    breaklines=true, 
    breakindent=0pt, 
    numberstyle=\footnotesize,
    caption={The primary terminating condition for a while loop is removed in \bziputil with \blade, highlighted in red. This can cause additional possibilities for infinite loops.},
    label={fig:removing_loop_break}
]    
while (1) {
while_continue: /* CIL Label */;
 @@ //primary break condition @@  
  @ if (!(i < 256)) {
        goto while_break;
    } @
    if (s->inUse[i]) {
        s->seqToUnseq[s->nInUse] = (UChar)i;
        (s->nInUse)++;
    }
    i++;
}
\end{lstlisting}

\begin{lstlisting}[
    basicstyle=\ttfamily\footnotesize, 
    linewidth=\linewidth, 
    breaklines=true, 
    breakindent=0pt, 
    caption={\newtext{Retained reference file functionality in \dateutil debloated using \cova (k=30), introducing vulnerability to access file modification times.}},
    label={fig:date-conservative-covA}
]
  if (tmp___17 != 0) {
    tmp___16 = __errno_location();
    error(1, *tmp___16, "%s", reference);
  }
  when = get_stat_mtime((struct stat const *)(&refstats)); @@//retrieve file modification time@@
\end{lstlisting}

\begin{lstlisting}[
    basicstyle=\ttfamily\footnotesize, 
    linewidth=\linewidth, 
    breaklines=true, 
    breakindent=0pt, 
    numberstyle=\footnotesize,
    caption={Code snippets that are completely removed by \chisel when debloating \rmutil on the \texttt{-rf} flags are highlighted. The figure illustrates the security and stopping condition checks at the beginning of the \emph{fts\_read} function that ensure the program returns before entering critical file tree search code in case of any errors.},
    label={fig:rm-fts-read-securitycheck}
]
@@// If no entry initialized for FTS traversal.@@
if (sp->fts_cur == ((void *)0)) {
  return ((FTSENT *)((void *)0));
} else {
  if (sp->fts_options & 8192) {
@@// Stop state flag set if traversal failure. @@
    return ((FTSENT *)((void *)0)); } }
\end{lstlisting}

\begin{lstlisting}[
     basicstyle=\ttfamily\footnotesize, 
    linewidth=\linewidth, 
    breaklines=true, 
    breakindent=0pt, 
    numberstyle=\footnotesize,
    caption={Code snippet removed by \blade from the \rmutil program. \blade removes this code snippet and runs the test cases to determine whether the code snippet should be kept or removed. The execution of the code file without this code snippet results in the removal of the \texttt{bin} folder in the Docker container.},
    label={rm_bin_deletion_example1}
]
if (!(sp->fts_options & 32)) {
@@// If first char dir is '.' (ASCII 46)@@
 if ((int)dp->d_name[0] == 46) { 
@@// If second character is null (dp->d_name = '.')@@
  if (!dp->d_name[1]) {
    goto __Cont; @@//Skip the the directory @@
  } else {
@@// If second character is dot '.' (ASCII 46)@@
     if ((int)dp->d_name[1] == 46) {
@@//if d_name[2] is null, (dp->d_name = '..')@@
      if (!dp->d_name[2]) {
          goto __Cont; @@// skip the directory @@
\end{lstlisting}

\begin{lstlisting}[
    basicstyle=\ttfamily\footnotesize, 
    linewidth=\linewidth, 
    breaklines=true, 
    breakindent=0pt, 
    numberstyle=\footnotesize,
    caption={Decompiled \texttt{make\_dir} function from the RAZOR-debloated binary of \mkdirutil. Error-handling, logging, and safety conditions surrounding the \texttt{mkdir} system call have been removed.},
    label={fig:razor-make-dir}
]
undefined4 make_dir(undefined8 param_1, ...){
    int iVar1;
    iVar1 = mkdir(param_1,param_3);
    if ((iVar1 == 0) == 0) { halt_baddata(); }
    if (param_4 != (uint *)0x0) {
    *param_4 = (uint)(iVar1 == 0);
    return 0;
    }
    // WARNING: Bad instruction ...
    halt_baddata();
}
\end{lstlisting}

\begin{lstlisting}[
    basicstyle=\ttfamily\footnotesize, 
    linewidth=\linewidth, 
    breaklines=true, 
    breakindent=0pt, 
    numberstyle=\footnotesize,
    caption={Lines of code highlighted in red represent the code removed by \blade in \dateutil. This is the only instance where the variable \emph{tmp\_\_\_2} is assigned before being returned, propagating garbage values further.},
    label={fig:uninitialized_variable}
]
static int to_hour(long hours, int meridian) { ...
  case_1:
    if (0L < hours) {
      ...
    } else {
    _L___0:@
      if (hours == 12L) {
        tmp___2 = 12;
      } else {
        tmp___2 = -1;
      }@
      tmp___3 = (long)tmp___2;
    }
    return ((int)tmp___3); } 
\end{lstlisting}

\begin{lstlisting}[
    basicstyle=\ttfamily\footnotesize, 
    linewidth=\linewidth, 
    breaklines=true, 
    breakindent=0pt, 
    numberstyle=\footnotesize,
    caption={Code snippets taken from \mkdirutil, debloated on \blade for the \emph{-m} flag. The removal of a pointer's initialization in a function leads to corruption of data within the program.},
    label={fig:mkdir-memory-leak}
]
strtol_error xstrtoul(char const *s, ...) {
   q = s;@
   p = &t_ptr;@ @@// this initialization was removed @@ @@
// uninitialised pointer p passed to strtoul, leading to corruption of argv[3] in the stack. @@
  tmp = strtoul(..., p, strtol_base);
...
int main(int argc, char **argv) {
...
@@ // mode_compile calls xstrtoul @@
tmp___2 = mode_compile(...);
change = tmp___2;
newmode = mode_adjust(newmode, ...);
 if (!(optind < argc)) { }
 {
@@ // argv[3] i.e the directory name has gotten corrupted at this point. @@
   dir___0 = (char const *)*(argv + optind);
   fail = make_dir(dir___0, ...)
...
\end{lstlisting}

\bibliographystyle{IEEEtran}
\bibliography{custom}

@misc{SIR,
  howpublished = {\url{https://sir.csc.ncsu.edu/portal/index.php}},
    title={Software-artifact Infrastructure Repository}, 
  note = {Accessed 2025-01-22},
}

@misc{dockerslim, 
  title = {Dockerslim}, 
  howpublished = {\url{https://github.com/slimtoolkit/slim}}
}

@inproceedings{Blade,
  author       = {Muaz Ali and
                  Rumaisa Habib and
                  Ashish Gehani and
                  Sazzadur Rahaman and
                  Zartash Afzal Uzmi},
  title        = {{BLADE:} Towards Scalable Source Code Debloating},
  booktitle    = {{IEEE} Secure Development Conference, SecDev 2023, Atlanta, GA, USA,
                  October 18-20, 2023},
  pages        = {75--87},
  publisher    = {{IEEE}},
  year         = {2023}
}

@inproceedings{DBLP:conf/feast/AlhanahnahBG24,
  author       = {Mohannad Alhanahnah and
                  Yazan Boshmaf and
                  Ashish Gehani},
  editor       = {Ryan Craven and
                  Matthew S. Mickelson},
  title        = {SoK: Software Debloating Landscape and Future Directions},
  booktitle    = {Proceedings of the 2024 Workshop on Forming an Ecosystem Around Software
                  Transformation, {FEAST} 2024, Salt Lake City, UT, USA, October 14-18,
                  2024},
  pages        = {11--18},
  publisher    = {{ACM}},
  year         = {2024},
  url          = {https://doi.org/10.1145/3689937.3695792},
  doi          = {10.1145/3689937.3695792},
  bibsource    = {dblp computer science bibliography, https://dblp.org}
}

@inproceedings{Chisel,
  author       = {Kihong Heo and
                  Woosuk Lee and
                  Pardis Pashakhanloo and
                  Mayur Naik},
  editor       = {David Lie and
                  Mohammad Mannan and
                  Michael Backes and
                  XiaoFeng Wang},
  title        = {Effective Program Debloating via Reinforcement Learning},
  booktitle    = {Proceedings of the 2018 {ACM} {SIGSAC} Conference on Computer and
                  Communications Security, {CCS} 2018, Toronto, ON, Canada, October
                  15-19, 2018},
  pages        = {380--394},
  publisher    = {{ACM}},
  year         = {2018},
}

@inproceedings{Cimplifier,
  author       = {Vaibhav Rastogi and
                  Drew Davidson and
                  Lorenzo De Carli and
                  Somesh Jha and
                  Patrick D. McDaniel},
  editor       = {Eric Bodden and
                  Wilhelm Sch{\"{a}}fer and
                  Arie van Deursen and
                  Andrea Zisman},
  title        = {Cimplifier: automatically debloating containers},
  booktitle    = {Proceedings of the 2017 11th Joint Meeting on Foundations of Software
                  Engineering, {ESEC/FSE} 2017, Paderborn, Germany, September 4-8, 2017},
  pages        = {476--486},
  publisher    = {{ACM}},
  year         = {2017},
  url          = {https://doi.org/10.1145/3106237.3106271},
  doi          = {10.1145/3106237.3106271},
  bibsource    = {dblp computer science bibliography, https://dblp.org}
}

@inproceedings{Piecewise,
  author    = {Anh Quach and
               Aravind Prakash and
               Lok{-}Kwong Yan},
  title     = {Debloating Software through Piece-Wise Compilation and Loading},
  booktitle = {USENIX Security},
  pages     = {869--886},
  year      = {2018}
}

@inproceedings{Cov,
  author       = {Qi Xin and
                  Qirun Zhang and
                  Alessandro Orso},
  title        = {Studying and Understanding the Tradeoffs Between Generality and Reduction
                  in Software Debloating},
  booktitle    = {37th {IEEE/ACM} International Conference on Automated Software Engineering,
                  {ASE} 2022, Rochester, MI, USA, October 10-14, 2022},
  pages        = {99:1--99:13},
  publisher    = {{ACM}},
  year         = {2022},
}

@inproceedings{razor,
  author       = {Chenxiong Qian and
                  Hong Hu and
                  Mansour Alharthi and
                  Simon Pak Ho Chung and
                  Taesoo Kim and
                  Wenke Lee},
  editor       = {Nadia Heninger and
                  Patrick Traynor},
  title        = {{RAZOR:} {A} Framework for Post-deployment Software Debloating},
  booktitle    = {28th {USENIX} Security Symposium, {USENIX} Security 2019, Santa Clara,
                  CA, USA, August 14-16, 2019},
  pages        = {1733--1750},
  publisher    = {{USENIX} Association},
  year         = {2019},
  url          = {https://www.usenix.org/conference/usenixsecurity19/presentation/qian},
  bibsource    = {dblp computer science bibliography, https://dblp.org}
}

@inproceedings{Creduce,
  author       = {John Regehr and
                  Yang Chen and
                  Pascal Cuoq and
                  Eric Eide and
                  Chucky Ellison and
                  Xuejun Yang},
  editor       = {Jan Vitek and
                  Haibo Lin and
                  Frank Tip},
  title        = {Test-case reduction for {C} compiler bugs},
  booktitle    = {{ACM} {SIGPLAN} Conference on Programming Language Design and Implementation,
                  {PLDI} '12, Beijing, China - June 11 - 16, 2012},
  pages        = {335--346},
  publisher    = {{ACM}},
  year         = {2012},
  url          = {https://doi.org/10.1145/2254064.2254104},
  doi          = {10.1145/2254064.2254104},
  bibsource    = {dblp computer science bibliography, https://dblp.org}
}

@inproceedings{Perses,
  author       = {Chengnian Sun and
                  Yuanbo Li and
                  Qirun Zhang and
                  Tianxiao Gu and
                  Zhendong Su},
  editor       = {Michel Chaudron and
                  Ivica Crnkovic and
                  Marsha Chechik and
                  Mark Harman},
  title        = {Perses: syntax-guided program reduction},
  booktitle    = {Proceedings of the 40th International Conference on Software Engineering,
                  {ICSE} 2018, Gothenburg, Sweden, May 27 - June 03, 2018},
  pages        = {361--371},
  publisher    = {{ACM}},
  year         = {2018},
  url          = {https://doi.org/10.1145/3180155.3180236},
  doi          = {10.1145/3180155.3180236},
  bibsource    = {dblp computer science bibliography, https://dblp.org}
}

@inproceedings{Domgrad,
  author       = {Qi Xin and
                  Myeongsoo Kim and
                  Qirun Zhang and
                  Alessandro Orso},
  title        = {Subdomain-Based Generality-Aware Debloating},
  booktitle    = {35th {IEEE/ACM} International Conference on Automated Software Engineering,
                  {ASE} 2020, Melbourne, Australia, September 21-25, 2020},
  pages        = {224--236},
  publisher    = {{IEEE}},
  year         = {2020},
  url          = {https://doi.org/10.1145/3324884.3416644},
  doi          = {10.1145/3324884.3416644},
  bibsource    = {dblp computer science bibliography, https://dblp.org}
}

@inproceedings{Debop,
  author       = {Qi Xin and
                  Myeongsoo Kim and
                  Qirun Zhang and
                  Alessandro Orso},
  editor       = {Gregg Rothermel and
                  Doo{-}Hwan Bae},
  title        = {Program debloating via stochastic optimization},
  booktitle    = {{ICSE-NIER} 2020: 42nd International Conference on Software Engineering,
                  New Ideas and Emerging Results, Seoul, South Korea, 27 June - 19 July,
                  2020},
  pages        = {65--68},
  publisher    = {{ACM}},
  year         = {2020},
  url          = {https://doi.org/10.1145/3377816.3381739},
  doi          = {10.1145/3377816.3381739},
  bibsource    = {dblp computer science bibliography, https://dblp.org}
}

@inproceedings {evaldebloatingtools,
author = {Michael D. Brown and Adam Meily and Brian Fairservice and Akshay Sood and Jonathan Dorn and Eric Kilmer and Ronald Eytchison},
title = {A Broad Comparative Evaluation of Software Debloating Tools},
booktitle = {33rd USENIX Security Symposium (USENIX Security 24)},
year = {2024},
isbn = {978-1-939133-44-1},
address = {Philadelphia, PA},
pages = {3927--3943},
url = {https://www.usenix.org/conference/usenixsecurity24/presentation/brown},
publisher = {USENIX Association},
month = aug
}

@article{cohenkappa,
    author = {Jacob Cohen},
    title ={A Coefficient of Agreement for Nominal Scales},
    journal = {Educational and Psychological Measurement},
    volume = {20},
    number = {1},
    pages = {37-46},
    year = {1960},
    doi = {10.1177/001316446002000104},
    URL = {   
        https://doi.org/10.1177/001316446002000104
    },
    eprint = { 
        https://doi.org/10.1177/001316446002000104
    }
}

@inproceedings{bloat-ref1,
  author       = {Guoqing Xu and
                  Nick Mitchell and
                  Matthew Arnold and
                  Atanas Rountev and
                  Gary Sevitsky},
  editor       = {Gruia{-}Catalin Roman and
                  Kevin J. Sullivan},
  title        = {Software bloat analysis: finding, removing, and preventing performance
                  problems in modern large-scale object-oriented applications},
  booktitle    = {Proceedings of the Workshop on Future of Software Engineering Research,
                  FoSER 2010, at the 18th {ACM} {SIGSOFT} International Symposium on
                  Foundations of Software Engineering, 2010, Santa Fe, NM, USA, November
                  7-11, 2010},
  pages        = {421--426},
  publisher    = {{ACM}},
  year         = {2010},
}

@inproceedings{bloat-ref2,
  author       = {Joanna McGrenere and
                  Gale Moore},
  editor       = {Sidney S. Fels and
                  Pierre Poulin},
  title        = {Are We All In the Same "Bloat"?},
  booktitle    = {Proceedings of the Graphics Interface 2000 Conference, May 15-17,
                  2000, Montr{\'{e}}al, Qu{\'{e}}bec, Canada},
  pages        = {187--196},
  publisher    = {Canadian Human-Computer Communications Society},
  year         = {2000},
}

@inproceedings{bloat-ref3,
  author       = {Babak Amin Azad and
                  Pierre Laperdrix and
                  Nick Nikiforakis},
  editor       = {Nadia Heninger and
                  Patrick Traynor},
  title        = {Less is More: Quantifying the Security Benefits of Debloating Web
                  Applications},
  booktitle    = {28th {USENIX} Security Symposium, {USENIX} Security 2019, Santa Clara,
                  CA, USA, August 14-16, 2019},
  pages        = {1697--1714},
  publisher    = {{USENIX} Association},
  year         = {2019},
}

@inproceedings{brown2024broadcomparativeevaluationsoftware,
  author       = {Michael D. Brown and
                  Adam Meily and
                  Brian Fairservice and
                  Akshay Sood and
                  Jonathan Dorn and
                  Eric Kilmer and
                  Ronald Eytchison},
  editor       = {Davide Balzarotti and
                  Wenyuan Xu},
  title        = {A Broad Comparative Evaluation of Software Debloating Tools},
  booktitle    = {33rd {USENIX} Security Symposium, {USENIX} Security 2024, Philadelphia,
                  PA, USA, August 14-16, 2024},
  publisher    = {{USENIX} Association},
  year         = {2024},
}

@inproceedings{container-debloaters,
  author       = {Muhammad Hassan and
                  Talha Tahir and
                  Muhammad Farrukh and
                  Abdullah Naveed and
                  Anas Naeem and
                  Fareed Zaffar and
                  Fahad Shaon and
                  Ashish Gehani and
                  Sazzadur Rahaman},
  title        = {Evaluating Container Debloaters},
  booktitle    = {{IEEE} Secure Development Conference, SecDev 2023, Atlanta, GA, USA,
                  October 18-20, 2023},
  pages        = {88--98},
  publisher    = {{IEEE}},
  year         = {2023},
}

@inproceedings{ghavamnia2020confine,
  title={Confine: Automated system call policy generation for container attack surface reduction},
  author={Ghavamnia, Seyedhamed and Palit, Tapti and Benameur, Azzedine and Polychronakis, Michalis},
  booktitle={International Symposium on Research in Attacks, Intrusions and Defenses (RAID)},
  year={2020}
}

@inproceedings{lei2017speaker,
  title={SPEAKER: Split-phase execution of application containers},
  author={Lei, Lingguang and Sun, Jianhua and Sun, Kun and Shenefiel, Chris and Ma, Rui and Wang, Yuewu and Li, Qi},
  booktitle={Detection of Intrusions and Malware, and Vulnerability Assessment: 14th International Conference, DIMVA 2017, Bonn, Germany, July 6-7, 2017, Proceedings 14},
  pages={230--251},
  year={2017},
  organization={Springer}
}

@inproceedings{debloatbench-sok-gs,
  title={SoK: A Tale of Reduction, Security, and Correctness-Evaluating Program Debloating Paradigms and Their Compositions},
  author={Ali, Muaz and Muzammil, Muhammad and Karim, Faraz and Naeem, Ayesha and Haroon, Rukhshan and Haris, Muhammad and Nadeem, Huzaifah and Sabir, Waseem and Shaon, Fahad and Zaffar, Fareed and others},
  booktitle={European Symposium on Research in Computer Security},
  pages={229--249},
  year={2023},
  organization={Springer}
}

@inproceedings{bloatref4,
author = {Bhattacharya, Suparna and Rajamani, Karthick and Gopinath, K. and Gupta, Manish},
title = {The interplay of software bloat, hardware energy proportionality and system bottlenecks},
year = {2011},
isbn = {9781450309813},
publisher = {Association for Computing Machinery},
address = {New York, NY, USA},
url = {https://doi.org/10.1145/2039252.2039253},
doi = {10.1145/2039252.2039253},
booktitle = {Proceedings of the 4th Workshop on Power-Aware Computing and Systems},
articleno = {1},
numpages = {5},
location = {Cascais, Portugal},
series = {HotPower '11}
}

@inproceedings{bloat-ref7,
author = {Xu, Guoqing and Arnold, Matthew and Mitchell, Nick and Rountev, Atanas and Sevitsky, Gary},
title = {Go with the flow: profiling copies to find runtime bloat},
year = {2009},
isbn = {9781605583921},
publisher = {Association for Computing Machinery},
address = {New York, NY, USA},
url = {https://doi.org/10.1145/1542476.1542523},
doi = {10.1145/1542476.1542523},
booktitle = {Proceedings of the 30th ACM SIGPLAN Conference on Programming Language Design and Implementation},
pages = {419–430},
numpages = {12},
location = {Dublin, Ireland},
series = {PLDI '09}
}

@inproceedings{trimmer,
  author    = {Hashim Sharif and
               Muhammad Abubakar and
               Ashish Gehani and
               Fareed Zaffar},
  title     = {{TRIMMER:} application specialization for code debloating},
  booktitle = {Proceedings of the 33rd {ACM/IEEE} International Conference on Automated
               Software Engineering, {ASE} 2018, Montpellier, France, September 3-7,
               2018},
  pages     = {329--339},
  year      = {2018}
}

@inproceedings{lmcas-arxiv,
  author       = {Mohannad Alhanahnah and
                  Rithik Jain and
                  Vaibhav Rastogi and
                  Somesh Jha and
                  Thomas W. Reps},
  title        = {Lightweight, Multi-Stage, Compiler-Assisted Application Specialization},
  booktitle    = {7th {IEEE} European Symposium on Security and Privacy, EuroS{\&}P
                  2022, Genoa, Italy, June 6-10, 2022},
  pages        = {251--269},
  publisher    = {{IEEE}},
  year         = {2022}
}

@inproceedings{Ancile,
  author       = {Priyam Biswas and
                  Nathan Burow and
                  Mathias Payer},
  editor       = {Anupam Joshi and
                  Barbara Carminati and
                  Rakesh M. Verma},
  title        = {Code Specialization through Dynamic Feature Observation},
  booktitle    = {{CODASPY} '21: Eleventh {ACM} Conference on Data and Application Security
                  and Privacy, Virtual Event, USA, April 26-28, 2021},
  pages        = {257--268},
  publisher    = {{ACM}},
  year         = {2021},
  url          = {https://doi.org/10.1145/3422337.3447844},
  doi          = {10.1145/3422337.3447844},
  bibsource    = {dblp computer science bibliography, https://dblp.org}
}

@inproceedings{occam,
author = {Malecha, Gregory and Gehani, Ashish and Shankar, Natarajan},
title = {Automated software winnowing},
year = {2015},
isbn = {9781450331968},
publisher = {Association for Computing Machinery},
address = {New York, NY, USA},
url = {https://doi.org/10.1145/2695664.2695751},
doi = {10.1145/2695664.2695751},
booktitle = {Proceedings of the 30th Annual ACM Symposium on Applied Computing},
pages = {1504–1511},
numpages = {8},
location = {Salamanca, Spain},
series = {SAC '15}
}

@inproceedings{Saffire,
  author       = {Shachee Mishra and
                  Michalis Polychronakis},
  title        = {Saffire: Context-sensitive Function Specialization against Code Reuse
                  Attacks},
  booktitle    = {{IEEE} European Symposium on Security and Privacy, EuroS{\&}P
                  2020, Genoa, Italy, September 7-11, 2020},
  pages        = {17--33},
  publisher    = {{IEEE}},
  year         = {2020},
  url          = {https://doi.org/10.1109/EuroSP48549.2020.00010},
  doi          = {10.1109/EUROSP48549.2020.00010},
  bibsource    = {dblp computer science bibliography, https://dblp.org}
}

\end{document}